\documentclass{jetpl}
\twocolumn
\title{Dimensional crossover in topological matter: Evolution of the multiple Dirac point in the layered system to the flat band on the surface}

\rtitle{Dimensional crossover in topological matter
}

\sodtitle{Dimensional crossover in topological matter: Evolution of the multiple Dirac point in the layered system to the flat band on the surface
}

\author{Tero T. Heikkil\"a $^{*}$\/\thanks{e-mail: tero.heikkila@tkk.fi}
and G.E. Volovik $^{*+}$
\/\thanks{e-mail: volovik@boojum.hut.fi}
}

\rauthor{T.T. Heikkil\"a, G.E.Volovik}

\sodauthor{Heikkil\"a, Volovik}

\address{
$^{*}$ Low Temperature Laboratory, Aalto University, School of Science and
Technology, P.O. Box 15100, FI-00076 AALTO, Finland
\\
$^+$ Landau Institute for Theoretical Physics RAS, Kosygina 2,
119334 Moscow, Russia
  }

\dates{November, 2010; preprint arXiv:1010.0393}{*}


\abstract{We consider the dimensional crossover in the topological matter, which involves the transformation of different types of topologically protected zeroes in the fermionic spectrum. In the considered case, the multiple Dirac (Fermi) point in quasi 2-dimensional system evolves into the flat band on the surface of the 3-dimensional system when the number of atomic layers increases. This is accompanied by formation of the spiral nodal lines in the bulk. We also discuss the topological quantum phase transition at which the surface flat band shrinks and changes its chirality, while the nodal spiral changes its helicity. }

\begin{document}

\maketitle

\newcommand  {\version}{v15}

\section{Introduction}

Topological matter is characterized by nontrivial topology of the Green's function  in momentum space \cite{Volovik2003,Horava2005}. The topological objects in momentum space (zeroes in the spectrum of fermionic quasiparticles) in many respects are similar to the topological defects in real space, and are also described by different homotopy groups including the relative homotopy groups. In particular, the Fermi surface is the momentum-space analog of the vortex loop in superfluids/superconductors; the Fermi point (or Dirac point) corresponds to the real-space point defects, such as hedgehog (monopole)  in ferromagnets; the fully gapped topological matter is characterized by skyrmions in momentum space, which are analogs of non-singular objects -- textures; etc. 

Here we discuss the transformations of the topologically protected zeroes, which occur during the dimensional crossover from a 2-dimensional to a 3-dimensional system. We consider the dimensional  crossover which involves such topological objects as a  
nodal line in a 3D system; the flat band, which is an analog of the vortex sheet; 
 and the topologically protected  Dirac points with multiple  topological charge $|N|>1$ in quasi 2-dimensional substance,
 which are analogous to the multiply quantized vortex.

The  Fermi bands, where the energy vanishes in a finite region of the momentum space,  and thus zeroes in the fermionic spectrum have co-dimension 0, have been discussed in different systems. The flat band appears in the so-called fermionic condensate 
\cite{Khodel1990,NewClass,Volovik2007,Shaginyan2010}.  
Topologically protected flat band exists in the spectrum of fermion zero modes localized in the core of some vortices \cite{KopninSalomaa1991,Volovik1994,MisirpashaevVolovik1995}.
In particle physics, the Fermi band (called the Fermi ball) appears  in  a 2+1 dimensional nonrelativistic quantum field theory which is dual to a gravitational theory in the anti-de Sitter  background with a charged black hole
\cite{Sung-SikLee2009}.
The flat band has also been discussed on the surface of the multi-layered graphene \cite{Guinea2006} and on the surface of superconductors without inversion symmetry \cite{SchnyderRyu2010}.

The topologically protected  2-dimensional and 3-dimensional Dirac points with multiple  topological charge $N$ were considered both in  condensed matter \cite{VolovikKonyshev1988,Volovik2003,Volovik2007,Manes2007,Dietl-Piechon-Montambaux2008,Chong2008,Banerjee2009,Sun2010,Fu2010,HeikkilaVolovik2010} and for relativistic quantum vacua \cite{Volovik2001,Volovik2003,KlinkhamerVolovik2005,Volovik2007}. 
In the vicinity of the multiple Dirac point with topological charge $N$ the spectrum may have the form
$E^2 \propto p^{2N}$.  We consider the special model of the multilayered system discussed in \cite{HeikkilaVolovik2010}, where the topological charge $N$  of the Dirac point coincides with the number of layers. In this model,  when $N\rightarrow \infty$, the multiple Dirac point transforms to the flat band in the finite region of the two-dimensional momentum on the surface of the sample. The interior layers in the limit $N\rightarrow \infty$ transform  to the bulk state, which represents a semi-metal in which the nodal line (line of zeroes) forms a spiral. The projection of this spiral onto the edge layer produces the boundary of the flat band. The latter is similar to what occurs in superconductors without inversion symmetry, where the region of the flat band on the surface is also determined by the projection of the topological nodal line in the bulk on the corresponding surface \cite{SchnyderRyu2010}.  

\section{Flat band and spiral nodal line}

\begin{figure}[h]
\centering
\includegraphics[width=8cm]{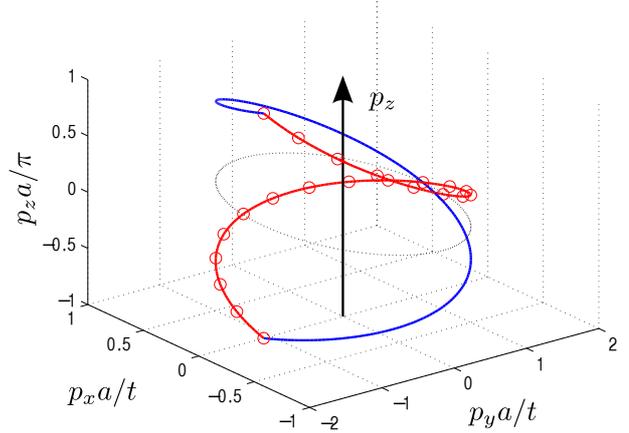}
\caption{Fig. 1. Fermi line for the case $t=|t_+|=10 |t_-|$ (red with circles) and $t=|t_-|=10 |t_+|$ (blue) along with their projection to the $p_z=0$ plane (dashed). This projection represents the boundary of the dispersionless flat band on the surface. In both cases $\phi_+=\phi_-=0$. Note that the helicity of the two lines is opposite, which gives the opposite signs of the invariant 
$N_1({\bf p}_\perp)=\pm 1$ in \eqref{InvariantForLine2}, and the opposite chiralities of the flat band.}
\label{fig:spiral}
\end{figure}

Let us first consider the model in Ref.~\cite{HeikkilaVolovik2010}, specified in 
Sec.~\ref{flatbandformation} 
 below, in the limit $N\rightarrow \infty$. The effective Hamiltonian in the 3-dimensional bulk system which emerges in the continuous limit $N\rightarrow \infty$ is the $2\times 2$ matrix
\begin{equation}
H=\begin{pmatrix}
0 & f\\
f^*& 0
\end{pmatrix}
~~,~~f=p_x-ip_y - t_+ e^{-ia p_z} -  t_-^* e^{ia p_z}  \,.
\label{ContinuousH}
\end{equation}
Here $t_+=|t_+|e^{i\phi_+}$ and $t_-=|t_- | e^{i\phi_-}$ are the hopping matrix elements between the layers and $a$ is the interlayer distance. The hopping matrix element proportional to $\sigma^{+(-)}$ is $t_{+(-)}$. 
The energy spectrum of the bulk system 
\begin{equation}
\begin{split}
&E^2= [p_x - |t_+| \cos  (ap_z-\phi_+)-|t_-|\cos(a p_z-\phi_-)]^2 
\\
&+  [p_y +| t_+| \sin  (ap_z-\phi_+)-|t_-|\sin(a p_z -\phi_-)]^2 \,,
\end{split}
\label{ContinuousE}
\end{equation}
 has zeroes on the line   (see Fig.~\ref{fig:spiral}):
 \begin{equation}
 \begin{split}
& p_x =| t_+| \cos  (ap_z-\phi_+)+|t_- |\cos(a p_z-\phi_-)\,,
\\
& p_y = |t_-| \sin(a p_z-\phi_-)-| t_+| \sin  (ap_z-\phi_+) \,.
\label{NodalLine}
\end{split}
\end{equation}
These zeroes are topologically protected by the topological invariant \cite{Volovik2007}
  \begin{equation} 
N_1=- {1\over 4\pi i} ~{\rm tr} ~\oint dl ~ \sigma_z H^{-1}\nabla_l H \,, 
\label{InvariantForLine}
\end{equation}
where the integral is along the loop around the nodal line in momentum space. 
The  winding number around the element of the nodal line is $N_1=1$.
For the interacting system the Hamiltonian matrix must be substituted by the inverse Green's function at zero frequency, $H\rightarrow G^{-1}(\omega=0, {\bf p})$, which plays the role of effective Hamiltonian, see also \cite{Gurarie2010}. 

%%%%%%%%%%%%%%%%%%%%%%%%%%%%%%%%%%%%%%%%%%%%%%%
%%%%%%%%%%%%%%%%%%%%%%%%%%%%%%%%%%%%%%%%%%%%%%%
%%%%%%%%%%%%%%%%%%%%%%%%%%%%%%%%%%%%%%%%%%%%%%%
%%%%%%%%%%%%%%%%%%%%%%%%%%%%%%%%%%%%%%%%%%%%%%%

 The same invariant can be written if the contour of integration is chosen parallel to $p_z$, i.e. at fixed ${\bf p}_\perp$. Due to periodic boundary conditions, the points $p_z=\pm \pi/a$ are equivalent and the contour of integrations forms the closed loop.
\begin{equation} 
N_1({\bf p}_\perp)=- {1\over 4\pi i} ~{\rm tr} ~\int_{-\pi/a}^{+\pi/a} dp_z ~ \sigma_z H^{-1}\nabla_{p_z} H\,, 
\label{InvariantForLine2}
\end{equation}
For interacting systems, this invariant can be represented in terms of the Green's  function expressed
via the 3D vector ${\bf g}(p_z,\omega)$ \cite{Volovik2007}:
\begin{equation}
G^{-1}(\omega,p_z)=ig_z(\omega,p_z) -  g_x(\omega,p_z) \sigma_x+   g_y(\omega,p_z)\sigma_y\,.
\label{IsingFermions2}
\end{equation}
In our model the components ${\bf g}(p_z,\omega)$ are: 
\begin{equation}
\begin{split}
& g_x(p_z,\omega)=p_x -| t_+| \cos  (ap_z-\phi_+)-|t_-| \cos(a p_z-\phi_-)
\\ 
& g_y(p_z,\omega)=p_y +| t_+| \sin(a p_z-\phi_+)-|t_-| \sin  (ap_z-\phi_-)\,,
\\
& g_z(p_z,\omega)=\omega \,.
\end{split}
\label{Components}
\end{equation}
Then the invariant \eqref{InvariantForLine2} becomes  \cite{Volovik2007}
\begin{equation}
N_1({\bf p}_\perp)={1\over 4\pi}\int_{-\pi/a}^{\pi/a} dp_z\int_{-\infty}^{\infty}  
d\omega~\hat{\bf g}\cdot
\left({\partial \hat{\bf g}\over\partial {p_z}} \times {\partial \hat{\bf
g}\over\partial {\omega}}\right)\,,
\label{2DInvariantIsing}
\end{equation}
where $\hat{\bf g}= {\bf g}/|{\bf g}|$.
It describes  the topological properties of the fully gapped 1D system, with $p_x$ and $p_y$ being the parameters of the system.

%%%%%%%%%%%%%%%%%%%%%%%%%%%%%%%%%%%%%%%%%%%%%%%
%%%%%%%%%%%%%%%%%%%%%%%%%%%%%%%%%%%%%%%%%%%%%%%
\begin{figure}[h]
\centering
\includegraphics[width=8cm]{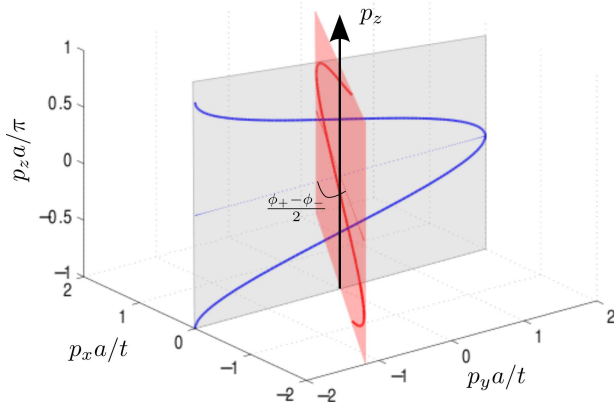}
\caption{Fig. 2. The Fermi line (nodal line) in the bulk at the topological quantum phase transition, which occurs at  $t=|t_+|=|t_-|$, for two values of the phase shift $\phi_- -\phi_+$. In this case the projection of the nodal line to the $p_z=0$ plane shrinks to the line segment with zero area because the Fermi line is flat: it is within the corresponding plane drawn on the figure. As a result the dispersionless flat band on the surface is absent at the transition, and has opposite chiralities on two sides of the transition.}
\label{fig:spiraltpequalstm}
\end{figure}
%%%%%%%%%%%%%%%%%%%%%%%%%%%%%%%%%%%%%%%%%%%%%%%
%%%%%%%%%%%%%%%%%%%%%%%%%%%%%%%%%%%%%%%%%%%%%%%

Let us first consider the case $t_-=0$. For  $t_-=0$, the nodal line in \eqref{NodalLine} forms a spiral,
the projection of this spiral on the plane $p_z={\rm const}$ being the circle 
 ${\bf p}_\perp^2\equiv p_x^2+p_y^2= |t_+|^2$ (the spiral survives for $t_- \neq 0$, but circle transforms to an ellipse, see Fig. \ref{fig:spiral} for the case $|t_-| < |t_+|$).
The topological charge in \eqref{InvariantForLine2} is $N_1({\bf p}_\perp)=1$ for momenta $|{\bf p}_\perp|<| t_+|$.  If the momentum ${\bf p}_\perp$ is considered as a parameter of the 1D system, then for $|{\bf p}_\perp|<| t_+|$  the system represents the 1D topological insulators. For $|{\bf p}_\perp|> | t_+|$ one has $N_1({\bf p}_\perp)=0$ and thus the non-topological 1D insulator. The line $|{\bf p}_\perp|=| t_+|$ marks the topological quantum phase transition between the topological and non-topological 1D insulators.

Topological invariant $N_1({\bf p}_\perp)$ in \eqref{InvariantForLine2} determines also the property of the surface bound states of the 1D system: the topological insulator must have the surface states with exactly zero energy. These states exist for any parameter within the circle  $|{\bf p}_\perp|=| t_+|$. This means that there is a flat band of states with exactly zero energy, $E(|{\bf p}_\perp|<| t_+|)=0$, which is protected by topology.
The bound states on the surface  of the system can be obtained directly from the Hamiltonian:
\begin{equation}
\begin{split}
&\hat H=\sigma_x (p_x - |t_+| \cos (a\hat p_z)) + \sigma_y (p_y + |t_+| \sin (a\hat p_z)) 
\\
&\hat p_z=-i\partial_z~~,~~ z<0\,.
\end{split}
\label{ContinuousHamilton}
\end{equation}
We assumed that the system occupies the half-space $z<0$ with the boundary at $z=0$, and made rotation in  $(p_x,p_y)$ plane to remove the phase $\phi_+$ of the hopping element $t_+$.
This Hamiltonian has the bound state with exactly zero energy, $E({\bf p}_\perp)=0$,  for any $|{\bf p}_\perp|<|t_+|$, with the eigenfunction concentrated near the surface:
 \begin{equation}
\Psi \propto \left( \begin{array}{cc}
0\\
1
\end{array} \right)
(p_x-ip_y) \exp{\frac{z \ln(t_+/(p_x+i p_y))}{a}}~~,~~|{\bf p}_\perp|< |t_+| \,.
\label{SurfaceWaveFunction}
\end{equation}
 The normalizable wave functions with zero energy exist only  for ${\bf p}_\perp$  within the circle $|{\bf p}_\perp|\leq |t_+|$, i.e. the surface flat band  is bounded by the projection of the nodal spiral onto the surface. Such correspondence between the flat band on the surface and lines of zeroes in the bulk has been also found in Ref.  \cite{SchnyderRyu2010}.
 
Restoring the non-zero hopping element $t_-$, we find that for 
$|t_-|<|t_+|$ there is still the region of the momentum ${\bf p}_\perp$ for which the topological charge in \eqref{InvariantForLine2} is $N_1({\bf p}_\perp)=1$. However, the area of the projection of the nodal line on the surface (and thus the area of the flat band) is reduced. Finally at $|t_-|=|t_+|$, the nodal line becomes flat,  its projection on the surface shrinks to the line segment $p_y/p_x=\tan (\phi_+-\phi_-)/2$, and the flat band disappears (Fig. \ref{fig:spiraltpequalstm}). For $|t_-|>|t_+|$, the spiral appears again, but the helicity of the spiral changes sign together with the topological charge in \eqref{InvariantForLine2}, which becomes $N_1({\bf p}_\perp)=-1$. Thus the point $|t_-|=|t_+|$ marks the topological quantum phase transition, at which the flat band changes its orientation (or actually its chirality). At the transition line $|t_-|=|t_+|$ the flat band on the surface does not exist. The non-zero helicity of the nodal line at $|t_-|\neq |t_+|$ reflects the broken inversion symmetry of the system at $|t_-|\neq |t_+|$. Note that in Ref. \cite{SchnyderRyu2010}  the flat surface bands also appeared in systems without inversion symmetry.

\section{Formation of the flat band in multilayered system}
\label{flatbandformation}

 We consider  the discrete model with finite number $N$ of layers. It is described by the $2N\times 2N$ Hamiltonian with the nearest neighbor interaction between the layers in the form:
\begin{equation}
\begin{split}
&H_{ij}({\bf p}_\perp)=
\\
&\boldsymbol{\sigma}\cdot{\bf p}_\perp \delta_{ij} - (t_+ \sigma^+ +  t_- \sigma^- )  \delta_{i,j+1} -(t_+^*\sigma^-  + t_-^*\sigma^+)\delta_{i,j-1}
\\
& 1\leq i\leq N~~,~~{\bf p}_\perp=(p_x,p_y)\,.
\label{DiscreteHamiltonian} 
\end{split}
\end{equation}
In the continuous limit of infinite number of layers \eqref{DiscreteHamiltonian} transforms to \eqref{ContinuousH}.
For $t_-=0$  and $t_+\equiv t$, equation \eqref{DiscreteHamiltonian}  represents the particular case of the model  \cite{HeikkilaVolovik2010}, which exhibits the Dirac point with multiple topological charge equal to $N$.

%%%%%%%%%%%%%%%%%%%%%%%%%%%%%%%%%%%%%%%%%%%%%%%
%%%%%%%%%%%%%%%%%%%%%%%%%%%%%%%%%%%%%%%%%%%%%%%
\begin{figure}[h]
\centering
\includegraphics[width=8cm]{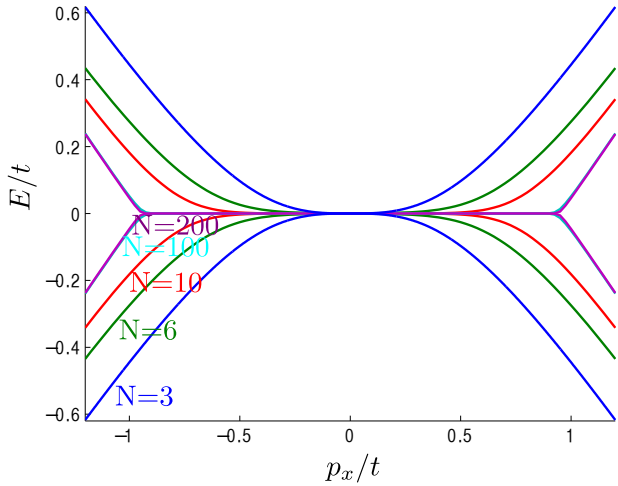}
\caption{Fig. 3. Formation of the surface flat band for parameters $t\equiv t_+  \in \Bbb{R}$, $t_-=0$. When  the number $N$ of layers increases, the  dispersionless band evolves from the   gapless branch of the spectrum, which has the form $E=\pm |{\bf p}_\perp|^N$ in the vicinity of multiple Dirac point. The spectrum is shown as a  function of $p_x$ for $p_y=0$ and finite value of $p_z$. The curves for $N=100$ and $N=200$ are almost on top of each other. Asymptotically  the spectrum $E=\pm |{\bf p}_\perp|^N$ transforms to the dispersionless band within the projection of the nodal line to the surface. }
    \label{spectrum_multiple_Dirac}
\end{figure}
%%%%%%%%%%%%%%%%%%%%%%%%%%%%%%%%%%%%%%%%%%%%%%%
%%%%%%%%%%%%%%%%%%%%%%%%%%%%%%%%%%%%%%%%%%%%%%%

Let us consider how this multiple Dirac point transforms to the flat band in the dimensional crossover, i.e. in limit $N\rightarrow\infty$ (see Fig. \ref{spectrum_multiple_Dirac}).  The Hamiltonian $H$ in  \eqref{DiscreteHamiltonian} with $t_-=0$  and $t_+\equiv t$ has two low-energy eigenstates with dispersion in the vicinity of multiple Dirac point  (at $|{\bf p}_\perp| \ll |t|$)
\begin{equation}
\epsilon/|t| \approx \pm (|{\bf p}_\perp|/|t|)^N.
\label{Nthorderdispersion}
\end{equation}
 Let us look for the eigenfunctions corresponding to the
dispersion \eqref{Nthorderdispersion}. At ${\bf p}_\perp=0$, the eigenfunctions are finite only in the first
and $N$'th layer,
\begin{equation}
\psi_{0\pm}=\psi({\bf p}_\perp=0)=\frac{1}{\sqrt{2}}[|\downarrow\rangle_1 \pm |\uparrow\rangle_N].
\end{equation}
That is: we get $H\psi_{0\pm}=0$. Now for $|{\bf p}_\perp| \ll |t|$ it
suffices to find a function $\delta \psi$ satisfying $H\delta \psi =
|{\bf p}_\perp|^N/|t|^{N-1}$. It turns out that the small parameter in
this expansion is dependent on the layer index, and the smallest
correction we have to include is $\eta \sim |{\bf p}_\perp|^N$. In this order,
we can expand the eigenvalue equation in $\eta$:
\begin{equation}
\begin{split}
&H\psi_{p}=H(\psi_0+\eta \delta \psi_1 + \eta^2 \delta \psi_2 +
\cdots)\\&=\eta (H \delta \psi_1 + \eta H \delta \psi_2 + \cdots)\\
&=\alpha \eta \psi_0 + \eta^2 H \delta \psi_2 = |{\bf p}_\perp|^N/|t|^{N-1} \psi_p 
\\
&=\eta (\psi_0/|t|^{N-1} + \eta \delta \psi_1/|t|^{N-1} + \cdots),
\end{split}
\end{equation}
where $\eta \delta \psi_1 = \delta \psi$. To lowest order in $\eta$ it
is thus sufficient to find this $\delta \psi$. Moreover, to preserve
the normalization, we seek $\delta \psi$ such that it is orthogonal to
$\psi_0$. In this case, to the lowest order in $\eta$ the
normalization remains unaltered.

Let us hence write $\delta \psi$ in the form
\begin{equation}
\delta \psi = \sum_{n=1}^N (\alpha_{n\uparrow} |\uparrow\rangle_n +
\alpha_{n\downarrow}|\downarrow\rangle_n
\end{equation}
so that $\alpha_{1\downarrow}=\alpha_{N\uparrow}=0$. Now acting with
the Hamiltonian yields
\begin{equation}
\begin{split}
H\delta \psi =& \sum_{n=1}^N \bigg[(p_x-i
  p_y)\alpha_{n\uparrow}|\downarrow\rangle_n + (p_x+i p_y)
  \alpha_{n\downarrow}|\uparrow \rangle_n \\&- t
  \alpha_{n+1,\downarrow}(1-\delta_{nN})|\uparrow\rangle_n - t^*
  \alpha_{n-1,\uparrow} (1-\delta_{n1})|\downarrow\rangle_n \bigg]\\=& \pm
  \frac{(p_x^2+p_y^2)^{N/2}}{\sqrt{2}|t|^{N-1}}(|\downarrow\rangle_1
  \pm 
  |\uparrow \rangle_N). 
\end{split}
\end{equation}
 
%%%%%%%%%%%%%%%%%%%%%%%%%%%%%%%%%%%%%%%%%%%%%%%
%%%%%%%%%%%%%%%%%%%%%%%%%%%%%%%%%%%%%%%%%%%%%%%
\begin{figure}[h]
\centering
\includegraphics[width=8cm]{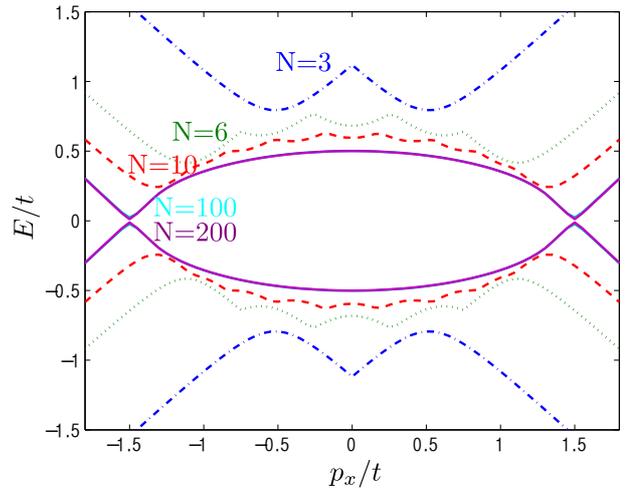}
\caption{Fig. 4. Formation of the nodal line from the evolution of the gapped branch of the spectrum of the multilayered system, when the number $N$ of layers increases.  
The spectrum is shown as function of $p_x$ for $p_y=0$ and finite value of $p_z$ for $t=t_+=2 t_- \in \Bbb{R}$. The curves for $N=100$ and $N=200$ lie almost on top of each other, indicating the bulk limit. Asymptotically the nodal line  $p_x = \pm (t_++t_-) \cos  (ap_z)$, $ p_y =  \pm (t_+-t_-) \sin  (ap_z)$ is formed (two points on this line are shown, which correspond to $p_y=0$).}
    \label{spectrum1}
\end{figure}
%%%%%%%%%%%%%%%%%%%%%%%%%%%%%%%%%%%%%%%%%%%%%%%
%%%%%%%%%%%%%%%%%%%%%%%%%%%%%%%%%%%%%%%%%%%%%%%

Considering this equation separately for each component results in
\begin{subequations}
\begin{align}
\alpha_{n\uparrow} &= \pm \frac{(-1)^{N-1}(p_x+ip_y)^{N/2}
  (p_x-ip_y)^{N/2-n}}{\sqrt{2}t^{(N-1)/2} (t^*)^{(N+1)/2-n}}\\
\alpha_{n\downarrow} &=
\frac{(-1)^{N-1}(p_x+ip_y)^{n-N/2}(p_x-ip_y)^{N/2}}{\sqrt{2}t^{n-(N-1)/2}(t^*)^{(N-1)/2}}.
\end{align}
\end{subequations}
The expressions are somewhat simpler for $p_y=0$:
\begin{subequations}
\begin{align}
\alpha_{n\uparrow} &= \pm \frac{(-1)^{N-1}}{\sqrt{2}}\left(\frac{p_x}{t^*}\right)^{N-n} \left(\frac{t^*}{t}\right)^{(N-1)/2}\\
\alpha_{n\downarrow} &=
\frac{(-1)^{N-1}}{\sqrt{2}} \left(\frac{p_x}{t}\right)^n \left(\frac{t}{t^*}\right)^{(N-1)/2}.
\end{align}
\end{subequations}
We hence find that the eigensolutions behave as $\sim |{\bf p}_\perp|^n$ for $n$
layers away from the surfaces, in agreement with  the wave function \eqref{SurfaceWaveFunction} for the surface flat band obtained in the continuous limit
\begin{equation}
\begin{split}
& \Psi \propto  \left( \begin{array}{cc}
0\\
1
\end{array} \right)(p_x-ip_y)\exp{\frac{z \ln(t/(p_x+ip_y))}{a}} 
\\& + \left( \begin{array}{cc}
1\\
0
\end{array} \right)(p_x+ip_y)\exp{\frac{-(L+z) \ln(t^*/(p_x-ip_y))}{a}} 
\\
& \overset{p_y = 0}{\approx }
 \left( \begin{array}{cc}
0\\
1
\end{array} \right)  \left(\frac{p_x}{t}\right)^{n}
+ \left( \begin{array}{cc}
1\\
0
\end{array} \right)  \left(\frac{p_x}{t^*}\right)^{N-n}
~~,~~|{\bf p}_\perp|<|t|\,.
\label{SurfaceWaveFunctionDiscrete}
\end{split}
\end{equation}

%%%%%%%%%%%%%%%%%%%%%%%%%%%%%%%%%%%%%%%%%%%%%%%
%%%%%%%%%%%%%%%%%%%%%%%%%%%%%%%%%%%%%%%%%%%%%%%
\begin{figure}[h]
\centering
\includegraphics[width=8cm]{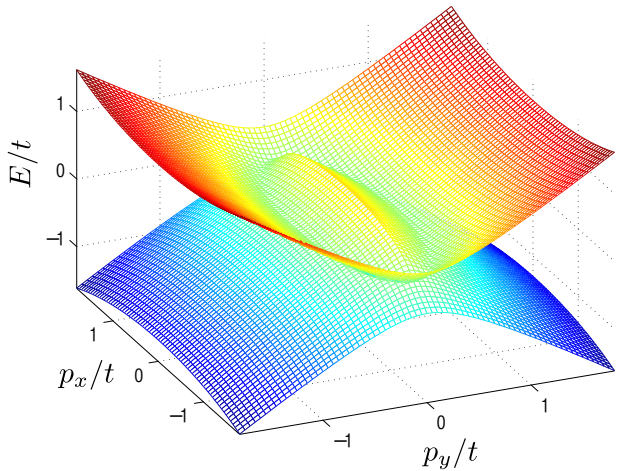}
\caption{Fig. 5. Gapped spectrum as a function of $p_x$ and $p_y$ for N=200 layers corresponding to the bulk limit. The nodal line forms at the projection of the Fermi line $p_x = \pm (t_++t_-) \cos  (ap_z)$, $ p_y =  \pm (t_+-t_-) \sin  (ap_z)$ to the $p_z$-plane.}
    \label{spectrummesh}
\end{figure}
%%%%%%%%%%%%%%%%%%%%%%%%%%%%%%%%%%%%%%%%%%%%%%%
%%%%%%%%%%%%%%%%%%%%%%%%%%%%%%%%%%%%%%%%%%%%%%%

%%%%%%%%%%%%%%%%%%%%%%%%%%%%%%%%%%%%%%%%%%%%%%%
%%%%%%%%%%%%%%%%%%%%%%%%%%%%%%%%%%%%%%%%%%%%%%%
\begin{figure}[h]
\centering
\includegraphics[width=8cm]{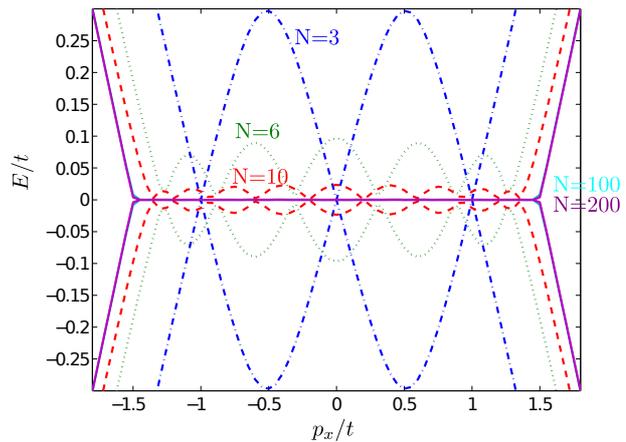}
\caption{Fig. 6. Evolution of the spectrum at $t=t_+=2 t_- \in \Bbb{R}$. The curves for $N=100$ and $N=200$ are almost on top of each other, indicating the bulk limit. Asymptotically the flat band is formed for $|p_x|<|t_+|+|t_-|$ and $|p_y|<|t_+|-|t_-|$. }
    \label{spectrum2}
\end{figure}
%%%%%%%%%%%%%%%%%%%%%%%%%%%%%%%%%%%%%%%%%%%%%%%
%%%%%%%%%%%%%%%%%%%%%%%%%%%%%%%%%%%%%%%%%%%%%%%

%%%%%%%%%%%%%%%%%%%%%%%%%%%%%%%%%%%%%%%%%%%%%%%
%%%%%%%%%%%%%%%%%%%%%%%%%%%%%%%%%%%%%%%%%%%%%%%
\begin{figure}[h]
\centering
\includegraphics[width=8cm]{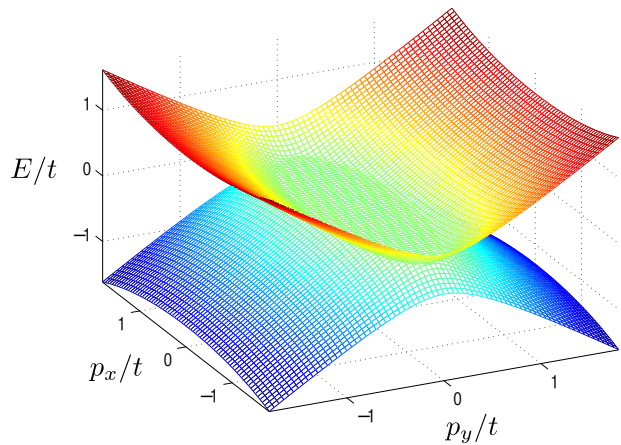}
\caption{Fig. 7. Energy of the surface states for different $p_x$ and $p_y$ for $N=200$ layers corresponding close to the bulk limit, calculated with the same parameters as in Fig.~\ref{spectrum2}. The flat band forms in the ellipse-shaped region $|p_x| < (t_++t_-)\cos(\theta)$, $|p_y| < (t_+-t_-)\sin(\theta)$, $\theta \in [0,2\pi]$. }
\label{lowenspectrummesh}
\end{figure}
%%%%%%%%%%%%%%%%%%%%%%%%%%%%%%%%%%%%%%%%%%%%%%%
%%%%%%%%%%%%%%%%%%%%%%%%%%%%%%%%%%%%%%%%%%%%%%%

We have considered the evolution of the multiple Dirac point with spectrum $E\sim \pm |{\bf p}_\perp|^N$ into the surface flat band when $N\rightarrow \infty$. 
However, the surface flat band survives when the coupling $t_-$ is added which splits the multiple Dirac point. The reason for the robustness of the flat band is the nodal line which is developed in the bulk (see Figs.~\ref{spectrum1} and \ref{spectrummesh}). As we discuss in the previous section, the nodal line supports  the topological stability of the flat band due to the bulk-surface correspondence.
Formation of the dispersionless flat band for  finite $|t_-|<|t_+|$ is shown in Figure~\ref{spectrum2}. The flat band has the ellipse-shaped region (Fig.~\ref{lowenspectrummesh}) whose boundary is the projection of the bulk nodal line on the surface.

\section{Conclusion}

We have considered the dimensional crossover in which the multiple Dirac (Fermi) point in quasi 2+1 system evolves into a flat band on the surface of the 3+1 system when the number of atomic layers increases. The formation of the surface flat band is a generic phenomenon which accompanies the formation of the nodal lines in bulk in the form of a spiral. We have also demonstrated a new type of a topological quantum phase transition, at which the flat band shrinks and changes its chirality. This transition is accompanied by the change of the helicity of the nodal line. The considered crossover is one of the numerous examples of the evolution of the topologically non-trivial quantum vacua, which are represented by the momentum-space topological objects. This example displays the ambivalent role of symmetry for these objects: the symmetry may support the topological charge of the object in momentum space,  or  may destroy the object.
In our case the time reversal symmetry supports the existence of the nodal line in bulk, and thus the flat band on the surface, while the space inversion symmetry kills the flat band.

Different scenarios of the dimensional crossover can be realized with cold atoms in optical lattices \cite{StanescuGalitskiSarma2010}. For our scenario we need the special stacking of graphene-like layers to have a spiraling nodal line. The spiral formed by zeroes in the energy spectrum has been discussed in Ref. \cite{McClure1969}  for rhombohedral graphite. The nodal line there is modified to the chain of the connected electron and hole Fermi pockets, so that   ``the Fermi surfaces resemble very long `link sausages' wound in a loose spiral''  \cite{McClure1969}. 

The dispersionless flat band also exists on the surface of the polar phase of triplet superfluid/superconductor. This superconductor obeys the time reversal and space inversion symmetry, and it has a line of zeroes in the form of a ring \cite{Volovik2003}. This ring gives rise to two surface flat bands with opposite chirality corresponding to two directions of spin. However, spin-orbit interaction may lead to the mutual annihilation of the flat bands.

More examples of dimensional crossover and exotic quantum phase transitions emerge when one considers the topology in the phase space, i.e. in the combined momentum-real $({\bf p},{\bf r})$ space \cite{GrinevichVolovik1988,Volovik2003}. This is appropriate in particular for the fermion zero modes localized on topological defects, which also may have dispersionless flat band \cite{KopninSalomaa1991,Volovik2003} and bulk-vortex correspondence 
\cite{TeoKane2010,SilaevVolovik2010}. The flat band inside the vortex core in 3-dimentional 
superfluids with Fermi (Dirac) points emerging due to the bulk-vortex correspondence is discussed in \cite{Volovik2010}.

Systems with topologically protected Fermi lines or Fermi points belong to the broad class of topological matter. As distinct from  topological insulators and superconductors/superfluids of the $^3$He-B type
\cite{HasanKane2010,QiZhang2010}, which belong to fully gapped topological matter, these are the gapless topological matter. However, it has the features which was earlier ascribed only to topological insulators, i.e. protected gapless states on the surface or inside the vortex core. If one or two components of the momentum ${\bf p}$ is fixed,   such as the projection $p_z$ of the momentum on the direction of the vortex axis 
(see accompanied paper \cite{Volovik2010}) or $|p_\perp| < |t|$ in our case,  
the system effectively behaves as one-dimensional and two-dimensional  topological insulator correspondingly. This is because for these parameters the system is fully gapped, while the effective 1D or 2D Hamiltonian has a non-trivial topology. Since the topological insulators cannot be adiabatically turned to a trivial insulator, this gives rise to the zero energy edge states in those intervals of parameters $p_z$ or  $|p_\perp| $, for which the topology is nontrivial. As a result, in  both cases one has the dispersionless spectrum with zero energy -- the flat band -- in the vortex core and on the surface of the system correspondingly.

This work is supported in part by the Academy of Finland, Centers of excellence program 2006--2011 and the European Research Council (Grant No. 240362-Heattronics).
It is our pleasure to thank N.B. Kopnin for helpful discussions.

\end{document}